\documentclass[10 pt,final,journal,letterpaper,oneside,twocolumn]{IEEEtran}
\usepackage{multicol}  
\usepackage{cite} 
\usepackage{multirow}
\usepackage[pdftex]{graphicx}
\usepackage[ruled]{algorithm2e}
\usepackage{algorithmic}
\usepackage[small]{caption}
\usepackage{float}
\usepackage[utf8]{inputenc}
\usepackage{amsmath, amsthm, amssymb}

\usepackage[english]{babel}
\usepackage{amsthm}
\usepackage{xspace}
\usepackage{fancyhdr}
\usepackage{amssymb,amsmath}
\usepackage{subfig}
\usepackage{textcomp} 
\usepackage{epstopdf}
\usepackage{bbding}
\usepackage{color}
\usepackage{pifont}

\begin{document}
%
\title{\LARGE Energy Efficiency Optimization for Backscatter Enhanced NOMA Cooperative V2X Communications under Imperfect CSI }

\author{ Wali Ullah Khan, Muhammad Ali Jamshed, Eva Lagunas, Symeon Chatzinotas,\\ Xingwang Li  and Bj\"orn Ottersten \thanks{The part of this work was submitted for presentation in IEEE VTC'2022-Spring, Helsinki, Finland, June 2022 \cite{conf}. 

Wali Ullah Khan, Eva Lagunas, Symeon Chatzinotas, and Bj\"orn Ottersten are with the Interdisciplinary Center for Security, Reliability
and Trust (SnT), University of Luxembourg, 1855 Luxembourg City, Luxembourg (Emails: \{waliullah.khan, eva.lagunas, symeon.chatzinotas,bjorn.ottersten\}@uni.lu). 

Muhammad Ali Jamshed is with James Watt School of Engineering, University of Glasgow, UK (email: muhammadali.jamshed@glasgow.ac.uk).

Xingwang Li is with the School of Physics and Electronic Information Engineering, Henan Polytechnic University, Jiaozuo, China (email: lixingwangbupt@gmail.com).

}}%

\maketitle

\begin{abstract}
Automotive-Industry 5.0 will use beyond fifth-generation (B5G) technologies to provide robust, computationally intelligent, and energy-efficient data sharing among various onboard sensors, vehicles, and other devices. Recently, ambient backscatter communications (AmBC) have gained significant interest in the research community for providing battery-free communications. AmBC can modulate useful data and reflect it towards near devices using the energy and frequency of existing RF signals. However, obtaining channel state information (CSI) for AmBC systems would be very challenging due to no pilot sequences and limited power. As one of the latest members of multiple access technology, non-orthogonal multiple access (NOMA) has emerged as a promising solution for connecting large-scale devices over the same spectral resources in B5G wireless networks. Under imperfect CSI, this paper provides a new optimization framework for energy-efficient transmission in AmBC enhanced NOMA cooperative vehicle-to-everything (V2X) networks. We simultaneously minimize the total transmit power of the V2X network by optimizing the power allocation at BS and reflection coefficient at backscatter sensors while guaranteeing the individual quality of services. The problem of total power minimization is formulated as non-convex optimization and coupled on multiple variables, making it complex and challenging. Therefore, we first decouple the original problem into two sub-problems and convert the nonlinear rate constraints into linear constraints. Then, we adopt the iterative sub-gradient method to obtain an efficient solution. For comparison, we also present a conventional NOMA cooperative V2X network without AmBC. Simulation results show the benefits of our proposed AmBC enhanced NOMA cooperative V2X network in terms of total achievable energy efficiency.
\end{abstract}

\begin{IEEEkeywords}
B5G, AmBC, NOMA, V2X, imperfect CSI.
\end{IEEEkeywords}

\IEEEpeerreviewmaketitle

\section{Introduction}
\IEEEPARstart{B}{eyond} fifth-generation (B5G) transportation systems will improve traffic efficiency, control, reliability, and passenger safety \cite{8248667,9583590,9521550}. In this regard, vehicle-to-everything (V2X) communication such as vehicle-to-vehicle, vehicle-to-pedestrian, vehicle-to-infrastructure, and vehicle-to-air is one of the promising technologies to make it possible \cite{liu20206g,9217500}. V2X is expected to support diverse quality of services (QoS), massive connectivity, offer very low latency, and energy-efficient communication \cite{wang2021green}. Recently, 3rd generation partnership project (3GPP) is also working on different V2X solutions regarding public safety \cite{9530506}. However, there still exist various research challenges in the current transportation systems that need to be investigated \cite{7328306}. One of the critical issues is the massive connectivity in B5G V2X networks which might not be possible through the existing orthogonal spectrum access resources, resulting in traffic and data congestion \cite{9288999}. Thus, new spectrum access technologies would be required to provide high spectral efficiency and support massive connectivity in B5G V2X networks. 

Recently, non-orthogonal multiple access (NOMA) using power domain multiplexing has emerged as one of the top air interface technology due to its high spectral efficiency and low latency \cite{8861078}. One of the main features of NOMA is to accommodate multiple devices over the same spectrum resource simultaneously \cite{liu2021application}. It can be achieved through two different techniques, i.e., superposition coding (SC) and successive interference cancellation (SIC). In particular, SC is used on the transmitter side to superimpose multiple signals over the same spectrum using different transmit power, and SIC is then applied on the receiver side to decode different signals \cite{9411862}. It is important to note that the power allocation for different signals over the same spectrum is done based on the channel conditions \cite{9409740}. A device with poor channel conditions will assign a high transmit power, and the one with good channel conditions will allocate a low transmit power \cite{9261140}. It guarantees the QoS of weak devices and also ensures fairness among different devices in the network \cite{ali2022fair1}.
The explosive growth of wireless devices in V2X communication and their diverse QoS requirements will consume huge energy \cite{wang2020climate,9210136,kwan2016review}. Recently, the research community in academia and industry is striving for various energy-efficient solutions to ease this critical situation \cite{8214104}. Some of the possible solutions are to utilize the solar energy, conversion of vehicles to electric and use the existing ambient RF signals for communication among different vehicles. 

Ambient backscatter communication (AmBC) has recently gained tremendous attention due to its ability to use the existing RF signals for modulating and reflecting useful information \cite{9261963}. The fundamental concept of AmBC is to design a device that can harvest energy from the existing RF signals to operate its circuit by modulating and reflecting own information \cite{khan2021integration,tran2020throughput}. This feature of the AmBC system make it the leading candidate technology towards battery-free communication in B5G era \cite{9261963}. Most of these devices are sensors that act as a tag by reflecting the existing RF signal towards near devices without exploiting any oscillatory circuity \cite{9348943}. From the last couple of years, the performance of these devices in various communication scenarios using conventional OMA technology has been extensively explored in the literature \cite{9363336,khan2021learning,9507551,9024401}.
\begin{table}[!t]
\fontsize{8.5}{11}\selectfont
\centering
\caption{List of notations and their definition}
\begin{tabular}{|c|c|} \hline 
\textbf{Notation} & \textbf{Definition}\\
\hline
$x$ & Superimposed signal of BS. \\ \hline
$m\in\{1,2\}$ & Represents RSU, i.e., $R_m$. \\ \hline
$P$ & Transmit power of BS. \\\hline
$\alpha_m$ & Power allocation coefficient of $R_m$. \\ \hline 
$x_m$ & Data symbol of $R_m$. \\ \hline 
$h_m$ & Denotes channel gain of $R_m$. \\ \hline 
$H_m$ & Denotes Rayleigh fading coefficient of $R_m$. \\ \hline 
$d_{m}$ & Distance between BS and $R_m$. \\ \hline 
$\zeta$ & Data symbol of $i$-th IE over $j$-th BF. \\ \hline 
$\hat h_{m}$ & Estimated channel gain of $R_m$. \\ \hline 
$\epsilon_m$ & Estimated channel error. \\ \hline
$\sigma_{\epsilon}^2$ & Variance of AWGN. \\ \hline  
$y_{m}$ & Received signal of $R_m$. \\ \hline 
$\varpi_{m}$ & Additive white Gaussian noise. \\ \hline 
${C}_{1},C_2$ & Data rate of $R_1$ and $R_2$. \\ \hline
$t_1,t_2$ & First time slot and second time slot. \\ \hline
$\gamma_1,\gamma_2$ & SINR of $R_1$ and $R_2$. \\ \hline   
$s_m$ & Superimposed signal of $R_m$. \\ \hline 
$Q_m$ & Transmit power of $R_m$ \\ \hline 
$i\in\{1,2\}$ & Represent vehicle of $R_m$, i.e., $V_{1,m}$. \\ \hline 
$\beta_{i,m}$ & Power allocation coefficient of $V_{i,m}$. \\ \hline 
$s_{i,m}$ & Data symbol of $V_{i,m}$. \\ \hline 
$B_m$ & Denotes backscatter tag in $R_m$. \\ \hline 
$z_m$ & Reflected data symbol of $B_m$. \\ \hline
$y_{i,m}$ & Received signal of $V_{i,m}$. \\ \hline 
$\hat h_{i,m}$ & Estimated channel gain of $V_{i,m}$. \\ \hline  
$\xi$ & Reflection coefficient of $B_m$. \\ \hline 
$\hat h_{i,m}^b$ & Estimated channel gain between $B_m$ of $V_{i,m}$. \\ \hline
$\hat h_{b,m}$ & Estimated channel gain between $B_m$ of $R_{m}$. \\ \hline
$\varpi_{i,m}$ & AWGN of $V_{i,m}$. \\ \hline
$C_{1,m},C_{2,m}$ & Data rate of $V_{1,m}$ of $V_{2,m}$. \\ \hline
$I_{i,m}^{m'}$ & Inter-RSU interference. \\ \hline
$\gamma_{1,m},\gamma_{2,m}$ & SINR of $V_{1,m}$ of $V_{2,m}$. \\ \hline
$\bar C$ & End-to-end data rate. \\ \hline
$\bar C_{sum}$ & End-to-end sum data rate. \\ \hline
$C_{min}$ & Minimum required data rate. \\ \hline
$P_{max}$ & Maximum power budget of BS. \\ \hline
$Q_{max}$ & Maximum power budget of $R_m$. \\ \hline
$L_1,L_2$ & Lagrangian function of (P1) and (P2). \\ \hline
$\psi_1,\psi_2$ & Lagrangian multipliers of $L_1$ in (P1). \\ \hline
$\lambda_1,\lambda_2$ & Lagrangian multipliers of $L_1$ in (P1). \\ \hline
$\eta_{1,m},\eta_{2,m},\mu_{i,m}$ & Lagrangian multipliers of $L_2$ in (P2). \\ \hline
$\zeta_m,\upsilon_m$ & Lagrangian multipliers of $L_2$ in (P2). \\ \hline
$\delta$ & Non-negative step size. \\ \hline
$t$ & Iteration index. \\ \hline
\end{tabular}
\label{tab1}
\end{table}

\subsection{Recent Literature}
Various research works have recently been investigated the performance of AmBC in NOMA networks. For example, Zhang {\em et al.} have proposed a AmBC enhanced downlink NOMA system to investigate the outage probability and ergodic sum rate \cite{zhang2019backscatter}. The work in \cite{zeb2019noma} have studied the system outage probability and throughput in AmBC enhanced NOMA communication. By integrating AmBC enhanced NOMA to air network, the authors of \cite{farajzadeh2019uav} have enhanced the successive bit rate while optimized the flight time and altitude of unmanned aerial vehicle. In another study, Chen {\em et al.} have investigated the expected sum capacity and outage probability of cooperative AmBC enhanced NOMA network \cite{chen2021backscatter}. Nazar {\em et al.} have provided AmBC enhanced NOMA framework to investigate the closed expression for bit error rate and bakscatter coverage \cite{nazar2021ber}. To analyse the secrecy of the system, Li {\em et al.} have computed the closed expressions of outage and intercept probability for AmBC enhanced NOMA network \cite{li2020secrecy}. Similar to \cite{zhang2019backscatter}, Raza {\em et al.} have proposed a massive machine type communication based on AmBC enhanced NOMA system and investigated the outage probability and ergodic rate \cite{raza2021noma}. To improve the reliability and security of maritime transmission system, Li {\em et al.} \cite{li2021cognitive} have calculated the outage and intercept probability in AmBC enhanced NOMA internet-of-vehicle network. Ding {\em et al.} have provided AmBC enhanced NOMA and wireless power transfer enhanced NOMA systems and investigate the outage probability and ergodic rate \cite{ding2021harvesting}. The authors of \cite{li2021hardware} have studied the impact of imperfect SIC, channel estimation error and residual hardware impairments on security and reliability performance of AmBC enhanced NOMA system. Further, the performance of AmBC enhanced NOMA with intelligent reflecting surfaces has also investigated by Le {\em et al.} in the form of outage probability, throughput and capacity \cite{le2021enabling}.

Researchers have also studied different resource allocation, and optimization frameworks in AmBC enhanced NOMA networks. For example, Khan {\em et al.} \cite{9543581} have presented a joint optimization framework for AmBC enhanced multi-cell NOMA network. They maximize the total achievable energy efficiency by optimizing the total transmit power of the base station (BS), users' power allocation, and the backscatter sensor's reflection coefficient under imperfect SIC decoding. To maximize the system throughput and fairness, the authors of \cite{liao2020resource} have jointly optimized the reflection coefficient, power and time allocation for full duplex AmBC enhanced NOMA network. In \cite{khan2021backscatter}, authors have maximized the spectral efficiency of AmBC enhanced NOMA V2X network by jointly optimize the transmit power of BS and roadside units (RSUs). Ding {\em et al.} have showed the advantages of using NOMA as a multiple access technique compared to other OMA techniques in AmBC network \cite{9497093}. The researchers of \cite{khan2021joint,khan2021backscatter11} have provided a new optimization framework to maximize the sum rate of AmBC enhanced NOMA network under perfect and imperfect SIC decoding. Moreover, Xu {\em et al.} \cite{9223730} have proposed a joint optimization framework for energy efficiency of AmBC enhanced NOMA network by allocating transmit power and designing reflection coefficient. To improve the secrecy in the presence of multiple non-colluding eavesdroppers, the works of \cite{khan2021secure,khan2020secure} have optimized resource allocation in single-cell and multi-cell AmBC enhanced NOMA network. Ding {\em et al.} \cite{9420716} have provided an new optimization framework to improve the uplink transmission rate while mitigate the interference between downlink and uplink of AmBC enhanced NOMA network. Furthermore, Ahmed {\em et al.} \cite{ahmed2021backscatter} investigated the energy efficiency optimization problem in multi-cell AmBC enhanced NOMA network under imperfect SIC decoding. In \cite{khan2021joint11}, the authors have presented a joint optimization framework to maximize the secrecy rate of AmBC enhanced NOMA network with multiple eavesdroppers. Of late, the research in \cite{khan2021ambient22} has explored a spectral efficiency optimization problem for multi-cell AmBC enhanced NOMA network. 

\subsection{Motivation and Contributions}
Despite extensive works on AmBC enhanced NOMA networks, the existing literature has not consider imperfect channel state information (CSI). In fact they consider perfect CSI in their proposed models, which is very challenging and hard in real systems. Besides that, most research works focus on non-vehicular and non-cooperative communications. To the best our knowledge, the problem of resource allocation for energy efficiency in AmBC enhanced NOMA cooperative V2X communication under imperfect CSI has not been considered previously. Motivated by this, we proposed a new optimization framework for energy efficiency in AmBC enhanced NOMA cooperative V2X network. In particular, we simultaneously optimize the transmit power of BS and RSUs while the reflection coefficient of backscatter sensor under imperfect CSI. We formulate the total transmit power minimization problem as non-convex and hard to solve jointly. Therefore, we first transform and decouple the original problem (denoted as $\mathcal P$) into two sub-problems, i.e., $\mathcal P1$ for first time slot (transmission from BS to RSUs) and $\mathcal P2$ for second time slot (transmission from RSUs to vehicles), respectively. Then we adopt iterative sub-gradient method to obtain an efficient solution. The main contributions of this work can be summarized as
\begin{enumerate}
    \item This paper proposes an energy-efficient optimization framework for AmBC enhanced NOMA cooperative V2X communication. In particular, we consider half duplex communication, where a BS transmit the superimposed signal to its serving RSUs in the first time slot following NOMA protocol. The RSUs act as relays, first decode the signal of BS and then forward to their associated vehicles in the second time slot. Meanwhile, the backscatter sensors also receive the signals of RSUs, modulate their own information and reflect towards vehicles. Thus, we simultaneously optimize the transmit power of BS and RSUs while the reflection coefficient of backscatter sensors under imperfect CSI. The objective is to minimize the total transmit power of V2X network while taking into account several practical constraints. 
    \item The problem of energy efficiency optimization is formulated as non-convex due to the imperfect CSI and co-channel interference. Moreover, our optimization problem is non-convex due to inter-cell interference and multiple optimization variables, making it very challenging to solve it directly. Therefore, an efficient way to solve the problem is to divide the original problem into two sub-problems. Then, we exploit the low-complex iterative sub-gradient method to obtain a sub-optimal yet efficient solution. In addition, we provide an algorithm to show different steps involve in the solution while also discussing its complexity.  
    \item Finally, we present the comprehensive numerical results based on Monte Carlo simulations. In the results, we characterize the efficiency of the proposed optimization framework based on sub-gradient algorithm. We demonstrate the impact of imperfect CSI on the total achievable energy efficiency of AmBC enhanced NOMA cooperative V2X network. We also show the effect of other key system parameters over the system performance such as total available transmit power, coverage of RSUs and circuit power consumption. In addition, we also provide the results of NOMA cooperative V2X communication without AmBC for comparison with the proposed framework.  
\end{enumerate}

The remainder of this paper is organized as follows. Section II discusses system model, various assumptions, and channel models at first time slot and second time slot. Section III explains and presents different constraints and problem formulation of total transmit power minimization. Section IV provides proposed solution based on iterative sub-gradient method. Section V presents simulation results and their discussion while Section VI concludes this paper with some future research directions. Various notations used in this paper are listed in Table I.  
\section{System and Channel Models}
\begin{figure*}[t]
\centering
\includegraphics [width=0.60\textwidth]{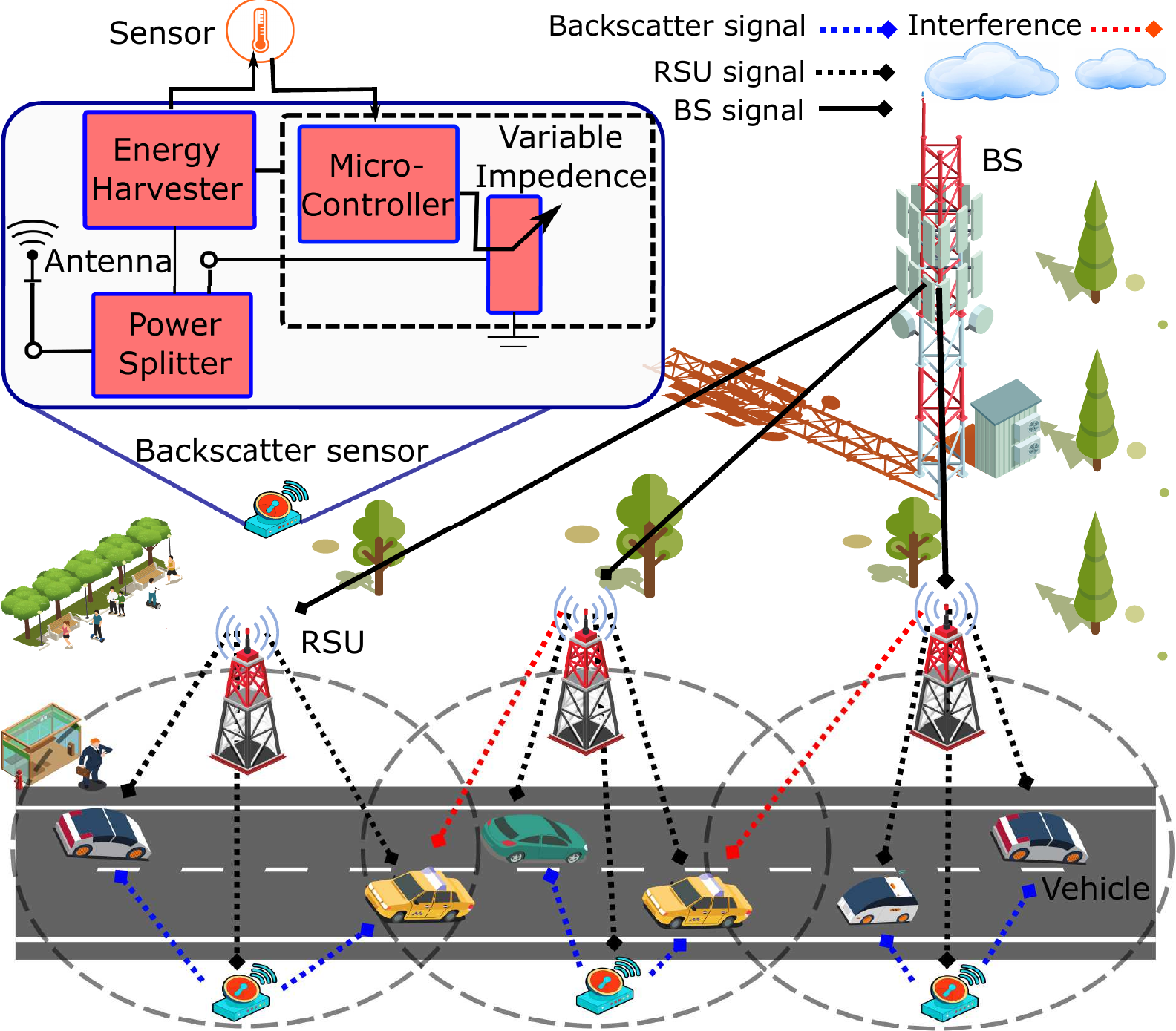}
\caption{System model of AmBC enhanced NOMA cooperative V2X communication.}
\label{blocky}
\end{figure*}
Consider an AmBC enhanced NOMA cooperative V2X communication, as illustrated in Figure \ref{blocky}. In the proposed model, we consider a half-duplex communication such that the data transmission from source to destination is completed in two time slots using NOMA protocol. More specifically, a BS transmits the superimposed signal to multiple RSUs in the first time slot. The RSUs first decode the signal of BS and then forward it to their associated vehicles in the second time slot. Meanwhile, the backscatter sensors also receive the signals of RSUs and reflect it toward vehicles by adding some useful information. In this work, we make the following assumptions: 1) The direct link from BS to vehicles and backscatter sensors is missing due to large distance and shadowing; 2) The BS, RSUs, backscatter sensors, and vehicles are using omnidirectional antenna; 3) The CSI is considered as perfect and imperfect; 4) The channels between different devices are independent and undergo Rayleigh fading. In the following, we discuss the channel models and signal to interference plus noise ratios at first and second time slot.

\subsection{First Time Slot (transmission from BS to RSUs)}
In the first time slot, transmission between BS and RSUs\footnote{To reduce the SIC decoding complexity and inter-RSU interference, we consider two RSUs associated with BS at any given time such that each RSU serves two vehicles.} takes place. A superimposed signal that BS transmits (denoted as $x$) can can be expressed as  
\begin{align}
x=\sum\limits_{m=1}^2\sqrt{P\alpha_m}x_m,
\end{align}
where $P$ is the transmit power of BS and $\alpha_m$ denotes the power allocation coefficient of RSU ( denoted as $R_m$), for $m\in\{1,2\}$. Moreover, $x_m$ represents the unit power signal of $R_m$. The channel from BS to $R_m$ can be modeled as
\begin{align}
h_m=H_m\times d_m^{\frac{-\zeta}{2}},
\end{align}
where $H_m$ is the Rayleigh fading coefficient, $d_m$ denotes the distance from BS to $R_m$ and $\zeta$ represents the pathloss exponent. In this work, we consider errors in channel estimation, hence the CSI is imperfect. Using the minimum mean square error (MMSE) model, the channel of $R_m$ from the BS is estimated as
\begin{align}
h_m=\hat h_m+\epsilon_m,
\end{align}
where $\hat h_m$ is the estimated channel gain of $h_m$ with variance $\sigma^2_{\hat h_m}$ and $\epsilon_m$ represents the estimated channel error with zero mean and $\sigma^2_{\epsilon_m}$ variance. For the convenience of discussion, the case of constant estimation error ($\sigma^2_{\epsilon_m}=\sigma^2_{\epsilon}$) for all channels is considered in this work. It is important to note that both $\hat h_m$ and $\epsilon_m$ are uncorrelated. The signal that $R_m$ receives from BS can be expressed as  
\begin{align}
y_m=\hat h_mx+\epsilon x+\varpi_m,\label{ym},
\end{align}
where $\varpi_m$ is the additive white Gaussian noise (AWGN) with zero mean and $\sigma^2$ variance. We assume that the channel gains of RSUs are arranged as $\hat h_{1}>\hat h_2$. Therefore, $R_1$ can apply SIC to decode its signal while $R_2$ cannot apply SIC and decode its signal by treating the signal of $R_1$ as a noise. Based on these observations, the data rate $C_1$ and $C_2$ can be written as $C_{1}=t_1W\log_2(1+\gamma_1)$ and $C_{2}=t_1W\log_2(1+\gamma_2)$, where $t_1$ shows the first transmission slot which should be equal to 1/2 and $W$ is the bandwidth. The terms $\gamma_1$ and $\gamma_2$ are the signal-to-interference-plus-noise-ratio (SINR) which can be stated as
\begin{align}
\gamma_1=\frac{|\hat h_1|^2P\alpha_1}{P\sigma^2_\epsilon(\alpha_1+\alpha_2)+\sigma^2}.
\end{align}
\begin{align}
\gamma_2=\frac{|\hat h_2|^2P\alpha_2}{|\hat h_2|^2P\alpha_1+P\sigma^2_\epsilon(\alpha_1+\alpha_2)+\sigma^2}.
\end{align}

\subsection{Second Time Slot (transmission from RSUs to vehicles)}
In this time slot, a transmission between RSUs and vehicles takes place. The RSUs first regenerate the superimposed signal and then forward it. The signal that $R_m$ transmits (denoted as $s_m$) to its associated vehicles can be written as 
\begin{align}
s_m=\sum\limits_{i=1}^2\sqrt{Q_m\beta_{i,m}}s_{i,m},
\end{align}
where $Q_m$ is the transmit power of $R_m$, $\beta_{i,m}$ represents the power allocation coefficient of the vehicle (denoted as $V_{i,m}$), for $i\in\{1,2\}$ and $s_{i,m}$ denotes the unit power signal of $V_{i,m}$. The channels used in this time slot are modeled and estimated similar to the first time slot. However, for the simplicity, we omit all the steps here and denote the channel from $R_m$ to $V_{i,m}$ as $\hat h_{i,m}$. Without loss of generality, we assume that the channel gains of $V_{1,m}$ is stronger than $V_{2,m}$, i.e., $|\hat h_{1,m}|^2>|\hat h_{2,m}|^2$.

During the transmission in the second time slot, a backscatter sensor in the geographical area of $R_m$ (stated as $B_{m}$) also receives the superimposed signal $s_m$ from $R_m$. $B_{m}$ first harvests energy from $s_m$, then modulate its own message $z_{m}$ and reflect it towards $V_{i,m}$, where $\mathbb E[|z_m|^2]=1$ and $\mathbb E[.]$ represents the expectation operation. Since we consider imperfect CSI, therefore the signal that $V_{i,m}$ receives from $R_m$ can be expressed as
\begin{align}
y_{i,m}&=\hat h_{i,m}s_m+\sqrt{\xi_m}\hat h^{b}_{i,m}\hat h_{b,m}s_mz_m\nonumber\\&+\epsilon s_m+\epsilon\sqrt{\xi_m}+\varpi_{i,m}, \label{y_{i,m}}
\end{align}
where $\xi_m$ is the reflection coefficient of $B_m$ and $\hat h^{b}_{i,m}$ denotes the channel gain between $B_m$ and $V_{i,m}$. Further, $\hat h_{b,m}$ represents the channel gain between $R_m$ and $B_m$ while $\varpi_{i,m}$ states the AWGN. Based on the received signal in (\ref{y_{i,m}}), the data rate of $V_{1,m}$ and $V_{2,m}$ cab be formulated as $C_{1,m}=t_2W\log_2(1+\gamma_{1,m})$ and $C_{2,m}=t_2W\log_2(1+\gamma_{2,m})$. Where $t_2$ represents the second transmission slot while $\gamma_{1,m}$ and $\gamma_{2,m}$ is the SINR of $V_{1,m}$ and $V_{2,m}$ as
\begin{align}
\gamma_{1,m}=\frac{Q_m\beta_{1,m}(|\hat h_{1,m}|^2+\xi_m|\hat h^{b}_{1,m}|^2|\hat h_{b,m}|^2)}{\sigma^2_\epsilon(Q_m(\beta_{1,m}+\beta_{2,m})+\xi_m)+I^{m'}_{1,m}+\sigma^2}.
\end{align}
\begin{align}
\gamma_{2,m}=\frac{Q_m\beta_{2,m}(|\hat h_{2,m}|^2+\xi_m|\hat h^{b}_{2,m}|^2|\hat h_{b,m}|^2)}{\varPi_{2,m}+\sigma^2_\epsilon(Q_m(\beta_{1,m}+\beta_{2,m})+\xi_m)+I^{m'}_{2,m}+\sigma^2}.
\end{align}
where in both equations, the term $\xi_m|\hat h_{1,m}^b|^2|\hat h_{b,m}|^2$ and $\xi_m|\hat h_{2,m}^b|^2|\hat h_{b,m}|^2$ in the nominator refer to the useful signals at vehicles received from the backscatter sensors. Furthermore, $\varPi_{2,m}=Q_m\beta_{1,m}(|\hat h_{2,m}|^2+\xi\hat h^{b}_{2,m}\hat h_{b,m})$ is the interference due to NOMA transmission, and $I^{m'}_{i,m}=|\hat h^{m'}_{i,m}|^2Q_{m'}$ is the inter-RSU interference due to the same channel reuse. Following decode-and-forward protocol at RSUs, the end-to-end rate can be calculated as 
\begin{align}
\bar C=\frac{1}{2}\text{min}\{C_m,C_{i,m}\}.
\end{align}
Accordingly, the sum rate of the proposed cooperative V2X network can be computed as
\begin{align}
\bar C_{sum}=\sum\limits_{m=1}^2\sum\limits_{i=1}^2\frac{1}{2}\text{min}\{C_m,C_{i,m}\}.
\end{align}

\section{Problem Formulation}
This work aims to provide energy-efficient communication in AmBC-enhanced NOMA cooperative V2X communication by optimizing various network resources, i.e., power allocation at BS and RSUs, while reflection coefficient at backscatter sensors. The total transmit power of the system can be written as
\begin{align}
 \sum\limits_{m=1}^2P\alpha_m+\sum\limits_{m=1}^2\sum\limits_{i=1}^2Q_m\beta_{i,m}.
\end{align}
The objective is to minimize the total power consumption while considering various practical constraints, which can be expressed as follows.
\begin{itemize}
\item\textit{Minimum data rate constraints:} To ensure the minimum data rate at both time slots, the minimum data rate should satisfy as
\begin{align}
\sum\limits_{m=1}^2C_m\geq C_{min}, 
\end{align}
\begin{align}
\sum\limits_{m=1}^2\sum\limits_{i=1}^2C_{i,m}\geq C_{min}, 
\end{align}
where $C_{min}$ is the minimum rate requirement for quality of services.
\item\textit{Power limit constraint:} To limit the total power at both time slots, the transmit power at BS and RSUs should satisfy as
\begin{align}
0\leq\sum\limits_{m=1}^2P\alpha_m\leq P_{max},
\end{align}
\begin{align}
0\leq\sum\limits_{m=1}^2\sum\limits_{i=1}^2Q_m\beta_{i,m}\leq Q_{max},
\end{align}
where $P_{max}$ and $Q_{max}$ denote the maximum power that BS and RSUs can transmit with.
\item\textit{Reflection coefficient constraint:} To control the reflection power of backscatter sensors in the second time slot, its reflection coefficient should be ranging between zero and one as
\begin{align}
0\leq\xi_m\leq1, m\in\{1,2\}.
\end{align}
\end{itemize} 
The formulation of the power minimization problem can also be represented as
\begin{alignat}{2}
\mathcal P &\ \underset{{\alpha_{m},\beta_{i,m},\xi_m}}{\text{min}} \bigg\{\sum\limits_{m=1}^2P\alpha_m+\sum\limits_{m=1}^2\sum\limits_{i=1}^2Q_m\beta_{i,m}\bigg\}\nonumber\\
s.t. &\  \mathcal A1: \sum\limits_{m=1}^2C_m\geq C_{min}, \nonumber\\
&\ \mathcal A2: \sum\limits_{i=1}^2C_{i,m}\geq C_{min}, m\in\{1,2\}, \nonumber\\
&\ \mathcal A3: \sum\limits_{m=1}^2P\alpha_m\leq P_{max},\nonumber\\
&\ \mathcal A4: \sum\limits_{m=1}^2\alpha_m\leq 1, \nonumber\\
&\ \mathcal A5: \sum\limits_{i=1}^2Q_m\beta_{i,m}\leq Q_{max}, m\in\{1,2\}, \nonumber\\
&\ \mathcal A6: \sum\limits_{i=1}^2\beta_{i,m}\leq 1, m\in\{1,2\},\nonumber\\
&\ \mathcal A7: 0\leq\xi_m\leq1, m\in\{1,2\},\label{19}
\end{alignat}
where the objective of $\mathcal P$ is to minimize the total transmit power of AmBC-enhanced NOMA cooperative V2X communication. Constraints $\mathcal A1$ and $\mathcal A2$ ensure the minimum data rate, where $C_{min}$ shows its threshold. Constraints $\mathcal A3$ and $\mathcal A5$ limit the transmit power of BS and RSUs, where $P_{max}$ and $Q_{max}$ are the maximum power that BS and RSU can transmit with at the given time. Constraints $\mathcal A4$ and $\mathcal A6$ describe the power allocation according to NOMA protocol. Constraint $\mathcal A7$ controls the reflection coefficient of backscatter sensors. We can see that the problem $\mathcal P$ is non-convex due to $\mathcal A1$ and $\mathcal A2$, making it very challenging to solve. Thus, we first transform and decouple it into two sub-problems and then employ the sub-gradient method to obtain an efficient solution.

\section{Proposed Energy-Efficient Solution}
The problem $\mathcal P$ is coupled on multiple variables which makes it very complex. Thus, we first transform and decouple it into two sub-problems, i.e., i) power allocation at BS; 2) power allocation at RSUs and reflection coefficient at backscatter sensors. Then, we adopt iterative sub-gradient method to obtain sub-optimal yet efficient solution \cite{boyd2004convex}. 

\subsection{Power Allocation for BS}
Here we optimize the power allocation of BS at first time slot for the fixed transmit power of RSUs and reflection coefficient of backscatter sensor at second time slot. By setting this, the original problem in (\ref{19}) can be reformulated as
\begin{alignat}{2}
\mathcal P1 &\ \underset{{\alpha_{1},\alpha_{2}}}{\text{min}}\ P(\alpha_1+\alpha_{2})\nonumber\\
s.t. &\  \mathcal A8:  |\hat h_1|^2P\alpha_1\geq(2^{C_{min}}-1)(P\sigma^2_\epsilon(\alpha_1+\alpha_2)+\sigma^2), \nonumber\\
& \ \mathcal A9: |\hat h_2|^2P\alpha_2\geq(2^{C_{min}}-1)\times\nonumber\\&\quad\quad\quad\ (|\hat h_2|^2P\alpha_1+P\sigma^2_\epsilon(\alpha_1+\alpha_2)+\sigma^2), \nonumber\\
& \ \mathcal A10: P(\alpha_1+\alpha_2)\leq P_{max},\nonumber\\
& \ \mathcal A11: \alpha_1+\alpha_2\leq 1,
\end{alignat}
where the objective of $\mathcal P1$ is to minimize the transmit power of BS. Constraint $\mathcal A8$ and constraint $\mathcal A9$ guarantee the minimum rate of $R_1$ and $R_2$, respectively. Constraint $\mathcal A10$ limits the transmit power of BS while constraint $(\mathcal A11)$ is the power allocation limit of $R_1$ and $R_2$. To solve problem (P1), we employ the sub-gradient method, in which we first define a Lagrangian function as
\begin{align}
L_1&(\psi_1,\psi_2,\lambda_1,\lambda_2)=P(\alpha_1+\alpha_2)+\psi_1((2^{C_{min}}-1)\nonumber\\&(P\sigma^2_\epsilon(\alpha_1+\alpha_2)+\sigma^2)-|\hat h_1|^2P\alpha_1)+\psi_2((2^{C_{min}}-1)\nonumber\\&(|\hat h_2|^2P\alpha_1+P\sigma^2_\epsilon(\alpha_1+\alpha_2)+\sigma^2)-|\hat h_2|^2P\alpha_2)\nonumber\\&+\lambda_1(P(\alpha_1+\alpha_2)-P_{max})+\lambda_2((\alpha_1+\alpha_2)-1),
\end{align}
where $\psi_1,\psi_2,\lambda_1, \lambda_2$ are the Lagrangian multipliers. After calculating the partial derivative, it can be written as
\begin{align}
\dfrac{\partial L_1}{\partial \alpha_1}& = P+\lambda_1 P+(2^{C_{min}}-1)\psi_2 (|\hat{h}_2|^2 P+P \sigma_{\epsilon}^2)\nonumber\\&+\psi_1 ((2^{C_{min}}-1) P \sigma_{\epsilon}^2-|\hat{h}_1|^2 P)+\lambda_2\alpha_1
\end{align}
\begin{align}
\dfrac{\partial L_1}{\partial \alpha_2}& = P+\lambda_1 P+(2^{C_{min}}-1) P \psi_1 \sigma_{\epsilon}^2\nonumber\\&+\psi_2 ((2^{C_{min}}-1) P \sigma_{\epsilon}^2-|\hat{h}_2|^2 P)+\lambda_2\alpha_2
\end{align}
Next we iteratively update the power allocation coefficients and Lagrangian multipliers as
\begin{align}
\alpha_1(t+1)&=\bigg(\alpha_1(t)- \delta(t) \dfrac{\partial L_1}{\partial \alpha_1}\bigg)^+
\end{align}
\begin{align}
\alpha_2(t+1)&=\bigg(\alpha_2(t)- \delta(t) \dfrac{\partial L_1}{\partial \alpha_2}\bigg)^+
\end{align}
\begin{align}
\psi_1(t+1) & =\big(\psi_1(t)-\delta(t)(|\hat h_1|^2P\alpha_1\nonumber\\&-(2^{C_{min}}-1)P\sigma_\epsilon^2(\alpha_1+\alpha_2)+\sigma^2)\big)^+
\end{align}
\begin{align}
\psi_2(t+1) & =\big(\psi_2(t)-\delta(t)(|\hat h_2|^2P\alpha_2-(2^{C_{min}}-1)\nonumber\\&(|\hat h_2|^2P\alpha_1+P\sigma_\epsilon^2(\alpha_1+\alpha_2))+\sigma^2)\big)^+
\end{align}
\begin{align}
\lambda_1(t+1)&=\big(\lambda_1(t)-\delta(t)(P_{max}-P(\alpha_1+\alpha_2))\big)^+
\end{align}
\begin{align}
\lambda_2(t+1)&=\big(\lambda_2(t)-\delta(t)(1-(\alpha_1+\alpha_2))\big)^+
\end{align}
where $t$ shows the iteration index and $\delta$ is the nonnegative step size. The above iterative process will continue until the required criterion is satisfied. 

\subsection{Power Allocation for RSUs and Reflection Coefficient for Backscatter Sensors}
Now we calculate the transmit power of RSUs and reflection coefficient of backscatter sensors at second time slot. Therefore, for a given transmit power at BS, the original problem in (\ref{19}) can be simplified as
\begin{alignat}{2}
\mathcal P2 & \underset{{\beta_{1,m},\beta_{2,m},\xi_{m}}}{\text{min}}\sum\limits_{m=1}^2Q_m(\beta_{1,m}+\beta_{2,m})\nonumber\\
s.t. &\  \mathcal A12:  Q_{m}\beta_{1,m}(|\hat h_{1,m}|^2+\varOmega_{1,m})\geq(2^{C_{min}}-1)\nonumber\\&\ (\sigma^2_\epsilon(Q_m(\beta_{1,m}+\beta_{2,m})+\xi_m)+I_{1,m}^{m'}+\sigma^2),m\in\{1,2\}, \nonumber\\
& \ \mathcal A13: Q_{m}\beta_{2,m}(|\hat h_{2,m}|^2+\varOmega_{2,m})\geq(2^{C_{min}}-1)(\varPi_{2,m}\nonumber\\&+\sigma^2_\epsilon(Q_m(\beta_{1,m}+\beta_{2,m})+\xi_m)+I_{2,m}^{m'}+\sigma^2),m\in\{1,2\}, \nonumber\\
& \ \mathcal A14: Q_m(\beta_{1,m}+\beta_{2,m})\leq Q_{max},m\in\{1,2\},\nonumber\\
& \ \mathcal A15: \beta_{1,m}+\beta_{2,m}\leq 1,m\in\{1,2\},\nonumber\\
& \ \mathcal A16: 0\leq\xi_m\leq 1,m\in\{1,2\},
\end{alignat}
where constraints $\mathcal A12$ and $\mathcal A13$ ensure the minimum rate of $V_{1,m}$ and $V_{2,m}$. Constraint $\mathcal A14$ and constraint $\mathcal A15$ control the transmit power of RSUs according to the NOMA protocol while $(A16)$ is the reflection coefficient constraint. Similar to $\mathcal P1$, here we also exploit the sub-gradient method. The Lagrangian function of $\mathcal P2$ can be defined as
\begin{align}
& L_2(\eta_{1,m},\eta_{2,m},\mu_{i,m},\zeta_m,\upsilon_{m})=Q_m(\beta_{1,m}+\beta_{2,m})\label{31}\\&+\eta_{1,m}((2^{C_{min}}-1)(\sigma^2_\epsilon(Q_m(\beta_{1,m}+\beta_{2,m})+\xi_m)\nonumber\\&+I_{1,m}^{m'}+\sigma^2)-Q_{m}\beta_{1,m}(|\hat h_{1,m}|^2+\xi_m|\hat h_{1,m}^{b}|^2|\hat h_{b,m}|^2))\nonumber\\&+\eta_{2,m}((2^{C_{min}}-1)(\varPi_{2,m}+\sigma^2_\epsilon(Q_m(\beta_{1,m}+\beta_{2,m})+\xi_m)\nonumber\\&+I_{2,m}^{m'}+\sigma^2)-Q_m\beta_{2,m}(|\hat h_{2,m}|^2+\xi_m|\hat h_{2,m}^{b}|^2|\hat h_{b,m}|^2))\nonumber\\&+\mu_{m}(Q_m(\beta_{1,m}+\beta_{2,m})-Q_{max})\nonumber\\&+\zeta_m((\beta_{1,m}+\beta_{2,m})-1)+\upsilon_m(\xi_m-1), m\in\{1,2\},\nonumber
\end{align}
where $\eta_{1,m},\eta_{2,m},\mu_{m},\zeta_m,\upsilon_m$ show the Lagrangian multipliers. Next we calculate the partial derivative of (\ref{31}) with respect to $\beta_{1,m}$, $\beta_{2,m}$, and $\xi_{m}$ which can be stated as
\begin{align}
\dfrac{\partial L_2}{\partial \beta_{1,m}}&= Q_m(1-|\hat{h}_{1,m}|^2 \eta_{1,m}-\xi_{m} |\hat{h}_{1,m}^b|^2 |\hat{h}_{b,m}|^2 \eta_{1,m}\nonumber\\&+(2^C_{min}-1)(\eta_{1,m}+\eta_{2})\sigma_{\epsilon}^2+ \mu_{m})
\end{align}
\begin{align}
\dfrac{\partial L_2}{\partial \beta_{2,m}}&= Q_m(1-|\hat{h}_{2,m}|^2 \eta_{2,m}-\xi_{m} |\hat{h}_{2,m}^b|^2 |\hat{h}_{b,m}|^2 \eta_{2,m}\nonumber\\&+(2^C_{min}-1)(\eta_{1,m}+\eta_{2,m})\sigma_{\epsilon}^2+ \mu_{m})
\end{align}
\begin{align}
\dfrac{\partial L_2}{\partial \xi_{m}}&=-\beta_{1,m}|\hat{h}_{1,m}^b|^2 |\hat{h}_{b,m}|^2 \eta_{1,m} Q_m-\beta_{2,m}|\hat{h}_{2,m}^b|^2 |\hat{h}_{b,m}|^2 \nonumber\\&\times\eta_{2,m} Q_m+(2^C_{min}-1)(\eta_{1}+\eta_{2,m})\sigma_{\epsilon}+\upsilon_m
\end{align}
To calculate efficient values, we iteratively update the optimization variables along with Lagrangian multipliers as
\begin{align}
&\beta_{1,m}(t+1)=\bigg(\beta_{1,m}(t)- \delta(t) \dfrac{\partial L_2}{\partial \beta_{1,m}}\bigg)^+
\end{align}
\begin{align}
&\beta_{2,m}(t+1)=\bigg(\beta_{2,m}(t)- \delta(t) \dfrac{\partial L_2}{\partial \beta_{2,m}}\bigg)^+
\end{align}
\begin{align}
&\xi_{m}(t+1)=\bigg(\xi_{m}(t)- \delta(t) \dfrac{\partial L_2}{\partial \xi_{m}}\bigg)^+
\end{align}
\begin{align}
\eta_{1,m}(t+1)&=\Big(\eta_{1,m}(t)- \delta(t)\big( Q_{m}\beta_{1,m}(|\hat h_{1,m}|^2\nonumber\\&+\xi_m|\hat h_{1,m}^{b}|^2|\hat h_{b,m}|^2-(2^{C_{min}}-1)(\sigma^2_\epsilon(Q_m\nonumber\\&\times(\beta_{1,m}+\beta_{2,m})+\xi_m)+I_{1,m}^{m'}+\sigma^2)\big)\Big)^+
\end{align}
\begin{align}
\eta_{2,m}(t+1)&=\Big(\eta_{2,m}(t)- \delta(t)\big( Q_{m}\beta_{2,m}(|\hat h_{2,m}|^2+\xi_m\nonumber\\&|\hat h_{2,m}^{b}|^2|\hat h_{b,m}|^2-(2^{C_{min}}-1)(\varPi_{2,m}+\sigma^2_\epsilon(Q_m\nonumber\\&\times(\beta_{1,m}+\beta_{2,m})+\xi_m)+I_{2,m}^{m'}+\sigma^2)\big)\Big)^+
\end{align}
\begin{align}
\mu_{m}(t+1)=\Big(\mu_{m}(t)-\delta(t)\big(Q_{max}-Q_m(\beta_{1,m}+\beta_{2,m})\big)\Big)^+
\end{align}
\begin{align}
\xi_{m}(t+1)=\Big(\xi_{m}(t)-\delta(t)\big(1-(\beta_{1,m}+\beta_{2,m})\big)\Big)^+
\end{align}
\begin{align}
\xi_{m}(t+1)=\Big(\xi_{m}(t)-\delta(t)\big(1-\xi_m\big)\Big)^+
\end{align}
where (35)-(42) are updated until the selection criterion is satisfied. In addition, a detailed steps involved in the proposed solution is also depicted in Algorithm 1.

\begin{algorithm}[t]

\caption{Energy Efficiency Optimization of V2X Network using Sub-gradient Method}
\begin{algorithmic}[1]

\STATE \textbf{INPUT} ($P_{\max}$, $\sigma_{\epsilon}$, $|\hat h_1|^2$, $|\hat h_2|^2$, $C_{\min}$, $|\hat h_{1,m}|^2$, $|\hat h_{2,m}|^2$, $\hat h^{b}_{1,m}|^2|$, $|\hat h^{b}_{2,m}|^2$, $|\hat h_{b,m}|^2$, and $\sigma^2$)
\STATE Set initial values of $\alpha_1=\alpha_2=0.5$;\\
\STATE Calculate initial NOMA interference using $|\hat h_2|^2P\alpha_1$;\\

\STATE \textbf{Stage 1:} Power allocation of hop 1 (First time slot transmission)

\REPEAT

\STATE Calculate $\alpha_1$, $\alpha_2$ and associated Lagrangian multipliers using (24)-(29) via iterative sub-gradient method;\\
\STATE Re-calculate $|\hat h_2|^2P\alpha_1$;
\UNTIL \textbf{convergence}

\STATE Calculate $Q_1=P\alpha_1$.

\STATE Calculate $Q_2=P\alpha_2$.

\STATE $Q_m= P(\alpha_1+\alpha_2)$.

\STATE \textbf{Stage 2:} Power allocation of hop 2 (Second time slot transmission)

\STATE For initial interference calculations set $Q_m$=$Q_{\max}/2$ $\forall m$;\\

\STATE Calculate initial inter-cell interference $I_{1,m}^{m'}$ and $I_{2,m}^{m'}$ $\forall m$;\\
\STATE Calculate initial NOMA interference $\varPi_{2,m}=Q_m\beta_{1,m}(|\hat h_{2,m}|^2+\xi\hat h^{b}_{2,m}\hat h_{b,m})$;\\
\REPEAT
\FOR{$m$ = 1 : $2$} 
\STATE Calculate $\beta_{1,m}$, $\beta_{2,m}$, $\xi_m$, and associated Lagrangian multipliers using (35)-(42) via iterative sub-gradient method;\\
\STATE Calculate $Q_{m'}$ using $\beta_{1,m}$, $\beta_{2,m}$;\\

\STATE Re-calculate $I_{1,m}^{m'}=|\hat h^{m'}_{1,m}|^2Q_{m'}$ and $I_{2,m}^{m'}=|\hat h^{m'}_{2,m}|^2Q_{m'}$;\\
\STATE Re-calculate $\varPi_{2,m}=Q_m\beta_{1,m}(|\hat h_{2,m}|^2+\xi\hat h^{b}_{2,m}\hat h_{b,m})$
\ENDFOR
\UNTIL \textbf{convergence}
\STATE \textbf{OUTPUT} $\alpha^*_1$, $\alpha^*_2$, $\beta^*_{1,1}$, $\beta^*_{1,2}$, $\beta^*_{2,1}$, $\beta^*_{2,2}$ , $\xi^*_1$, and  $\xi^*_2$.
\end{algorithmic}
 \label{AI-2}
\end{algorithm}
\subsection{Proposed Iterative Algorithm and Complexity Analysis}
Here we discuss the detailed steps of the proposed iterative sub-gradient method as depicted in Algorithm 1. First, the proposed algorithm takes the channel parameters, i.e.,  $|\hat h_1|^2$, $|\hat h_{2,m}|^2$, $\hat h^{b}_{1,m}|^2|$, etc., rate requirements $C_{\min}$ and power constraints $P_{\max}$, as an input parameters. Then energy efficiency is optimized in two stages, i.e., at the first time slot (stage 1) and second time slot (stage 2), respectively. In stage 1, the proposed algorithm optimizes the energy efficiency by allocating efficient power for both RSUs at BS, where the power allocation coefficients and Lagrangian multipliers are iteratively updated. In stage 2, for the given parameters at stage 1, the proposed algorithm optimizes the energy efficiency of each RSU by allocating efficient power to their associated vehicles. In this stage, our proposed algorithm also optimizes the reflection coefficients of backscatter sensors. Note that in both stages, the proposed algorithm ensures the QoS for RSUs and vehicles such that the value of $C_{\min}$ in both time slots is always satisfied. 

Here we also analyze the complexity analysis of our proposed iterative two-stage energy-efficient optimization algorithm. In this work, the computational complexity of the proposed algorithm can be calculated in terms of iterations required for the convergence of various optimization variables. In stage 1 and stage 2, the iterative sub-gradient method is used to optimize the power allocation of BS and RSUs while the reflection coefficients of backscatter sensors. Generally, the worst-case computational complexity of Algorithm 1 will be similar to the worst-case computational complexity of the sub-gradient method. Specifically, the worst-case computational complexity of the sub-gradient method can be computed as $\mathcal{O}(1/\Lambda^2)$, where $\Lambda$ is the number of iterations being used by the sub-gradient method to achieve the convergence. 

\section{Numerical Results and Discussion}
\begin{table}[!t]
\centering
\caption{Simulation parameters}
\begin{tabular}{|c||c|} 
\hline 
Parameter & Value  \\
\hline\hline
Total power budget $P_{tot}$ & 45 dBm \\\hline
Reflection coefficient  & $0\leq\xi\leq 1$ \\\hline
Channel type & Raleigh fading  \\\hline
RSU protocol & Decode-and-forward \\\hline
Imperfect CSI $(\sigma_\epsilon)$ & 0$\to$0.01\\\hline
Radius of BS & 50 meters\\\hline
Radius of RSU &  20 meters \\\hline
Channel realization & $10^3$ \\\hline
Minimum data rate $C_{\min}$ & 0.5 bps\\\hline
Pathloss exponent $(\zeta)$ & 4 \\\hline
Noise power density $\sigma^2$ & -170 dBm \\\hline
Bandwidth BW & 1 MHz \\\hline
Circuit power  & 5 dBm \\
\hline 
\end{tabular}
\end{table}
Here we present and discuss the numerical results of the proposed AmBC enhanced NOMA cooperative V2X communication based on Monte Carlo simulations. In addition, we also compare the results with the conventional NOMA cooperative V2X communication without AmBC. This work calculates the achievable energy efficiency as bits/Joule (Mb/J), which is the ratio between the total transmission rate of the network and the total power consumption, including the circuit power. Unless mentioned all the parameters used for simulation are provided in Table II.
\begin{figure}[t]
\centering
    \includegraphics[width=\columnwidth]{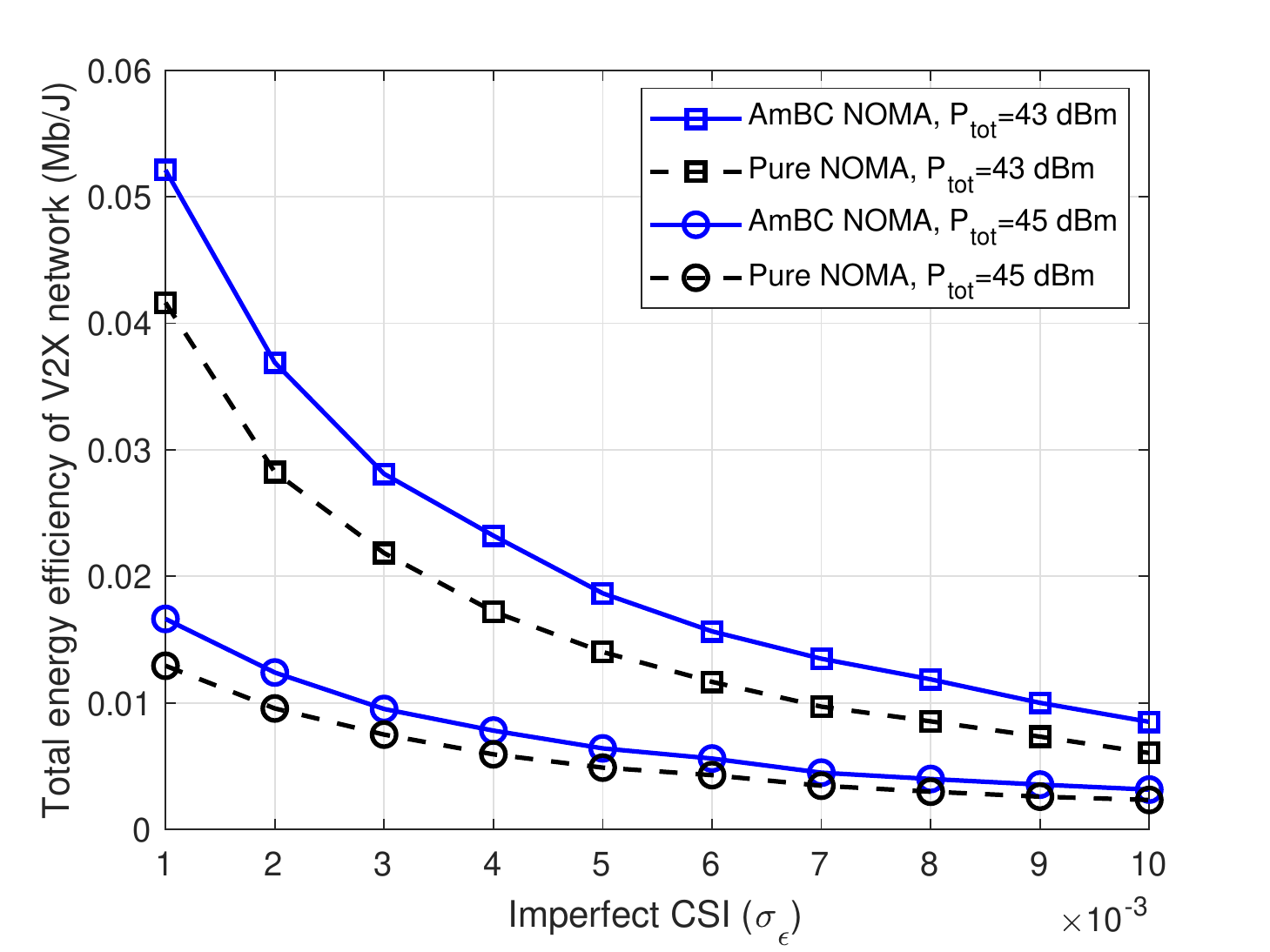}
    \caption{Comparison of system total energy efficiency versus imperfect CSI by varying total available power budget, i.e., $P_{\text{tot}}$.}
    \label{Fig1}
\end{figure}

It is important to study the impact of errors in the channel estimation on the system performance by plotting the total energy efficiency of cooperative V2X network versus the increasing values of imperfect CSI in Figure \ref{Fig1} for both AmBC enhanced NOMA and conventional NOMA systems. As expected, the achievable energy efficiency of the proposed AmBC enhanced V2X communication and the conventional cooperative VEX communication decreases as the values of imperfect CSI increases. This is because the interference due to imperfect CSI increases resulting decrease in the total transmission rate of both networks. However, the proposed AmBC enhanced NOMA cooperative communication achieves higher energy efficiency than the benchmark pure NOMA cooperative V2X communication. For example, when the total transmit power of both networks is $P_{tot}= 43$ dBm and the value of $\sigma_{\epsilon}$ is $10^{-3}$, the total energy efficiency of the proposed AmBC enhanced NOMA cooperative V2X communication is above 0.05 Mb/J. In comparison, the benchmark pure NOMA V2X communication can achieve only around 0.04 Mb/J. Moreover, similar trends can be seen when the total transmit power is $P_{tot}= 45$ dBm.  

\begin{figure}[t]
\centering
    \includegraphics[width=\columnwidth]{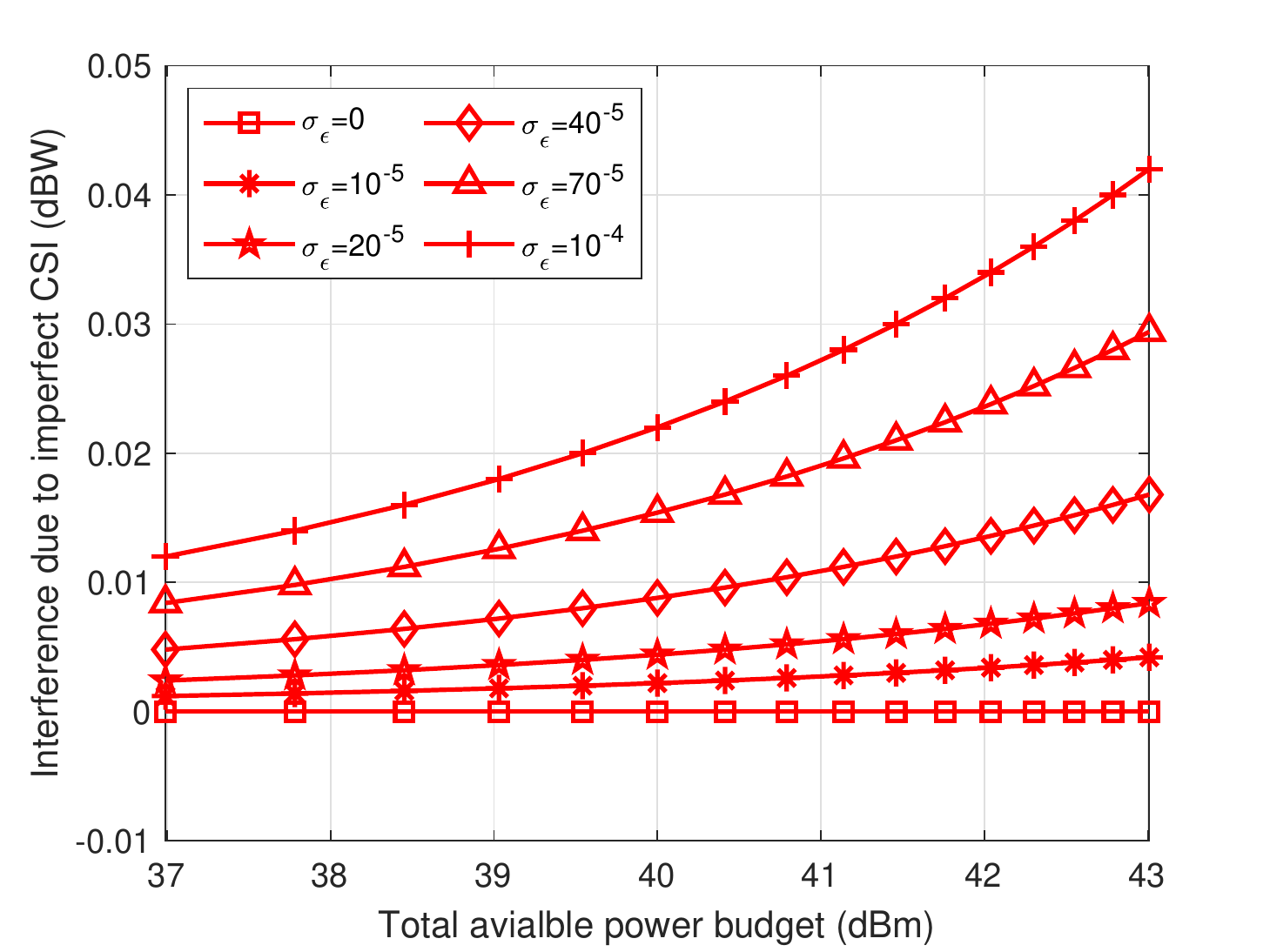}
    \caption{Total causes interference due to imperfect SCI versus total available power budget by varying $\sigma_{\epsilon}$.}
    \label{Fig2}
\end{figure}
Next, we investigate the amount of interference caused by various values of imperfect CSI. In this regard, Figure \ref{Fig2} shows the interference due to imperfect CSI versus total available power budget for different values of $\sigma_{\epsilon}$, i.e., $0, 10^{-5}, 20^{-5}, 40^{-5}, 70^{-5}, 10^{-4}$. Here we also show the result when the CSI is perfect. It can be evident that when the CSI is perfect, it causes zero interference for all values of the power budget. We can also observe that the interference is almost negligible for lower values of power budget and imperfect CSI. However, when the value of $\sigma_{\epsilon}$ increases, the interference due to imperfect CSI also increases. Furthermore, we can also see that the gap of interference among different curves increases as the total available power budget increase. It is because when the total available power budget increase, the system with high value of $\sigma_{\epsilon}$ produces high interference. This figure shows the importance of CSI and how much it can affect a network's performance.  

\begin{figure}[t]
\centering
    \includegraphics[width=\columnwidth]{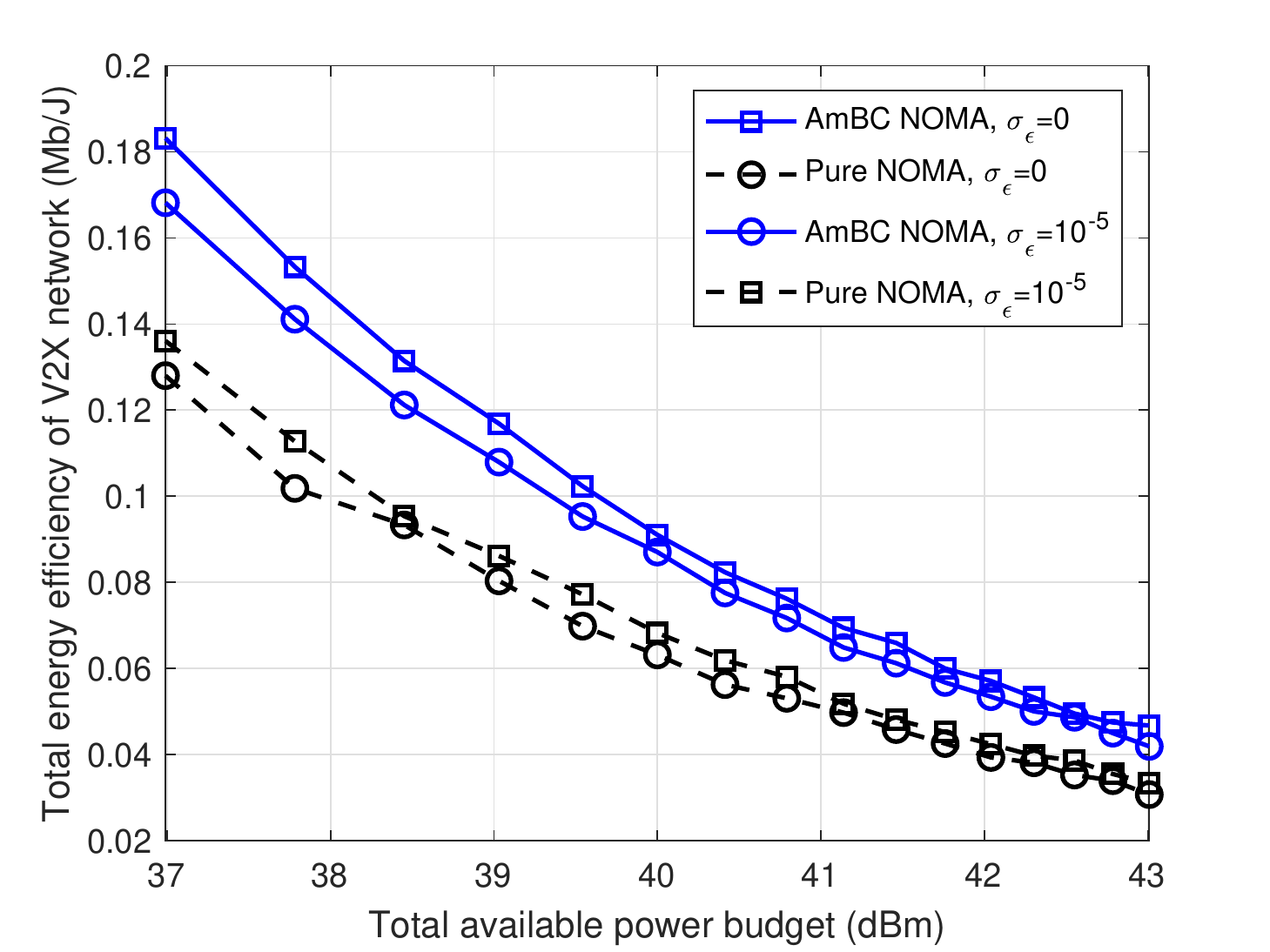}
    \caption{Comparison of system total energy efficiency versus the total available power budget by varying values of $\sigma_{\epsilon}$.}
    \label{Fig3}
\end{figure}
\begin{figure}[t]
\centering
    \includegraphics[width=\columnwidth]{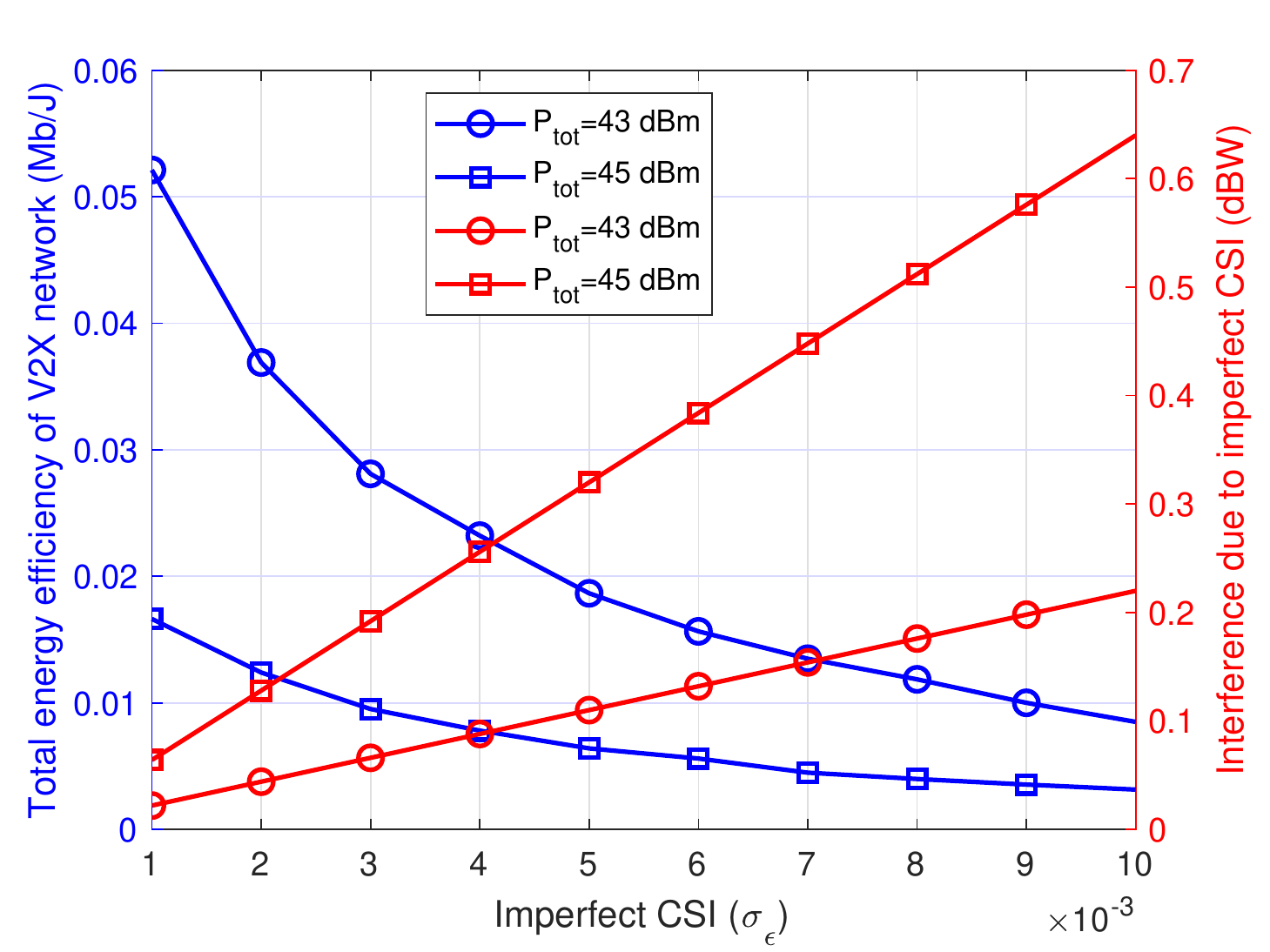}
    \caption{Trade-off between system total energy efficiency and interference due to imperfect CSI by varying $P_{tot}$.}
    \label{Fig4}
\end{figure}
Figure \ref{Fig3} shows the impact of transmit power on achievable energy efficiency of the system. For this figure, the values of imperfect CSI are set as $\sigma_{\epsilon}=0$ and $\sigma_{\epsilon}=10^{-5}$, respectively, and the transmit power varies from 37 dBm to 43 dBm. For both AmBC enhanced NOMA and pure NOMA, the network under the perfect CSI achieves high energy efficiency compared to the imperfect CSI case. This reflects the importance of estimating efficient CSI in the energy-efficient transmission of both networks. We note that the total energy efficiency of both V2X communications follows bell-shaped curves under it grows first with the network total available power budget until the saturating point and then drops with further increase in the power budget. However, the performance gap between the proposed and benchmark network shows the advantages of integrating AmBC in NOMA cooperative V2X Communication.

Figure \ref{Fig4} describes the trade-off between total energy efficiency versus the interference when varying the values of imperfect CSI. For this plot, we set the total available power budget of V2X network is $P_{tot}=43$ and $P_{tot}=45$ while the values of $\sigma_{\epsilon}$ ranges between $10^{-3}$ and $100^{-3}$, respectively. It is evident that the total energy efficiency of the proposed AmBC enhanced NOMA cooperative V2X communication decreases as the values of imperfect CSI, i.e., $\sigma_{\epsilon}$, increase. At the same time, the amount of interference is increases for the increasing values of $\sigma_{\epsilon}$. More specifically, for the lower values of $\sigma_{\epsilon}$, the proposed V2X communication cause low interference and achieves high energy efficiency. Another point is to note that when the system operates with a lower available power budget, i.e., 43, it causes low interference and achieves very high energy efficiency. 

\begin{figure}[t]
\centering
    \includegraphics[width=\columnwidth]{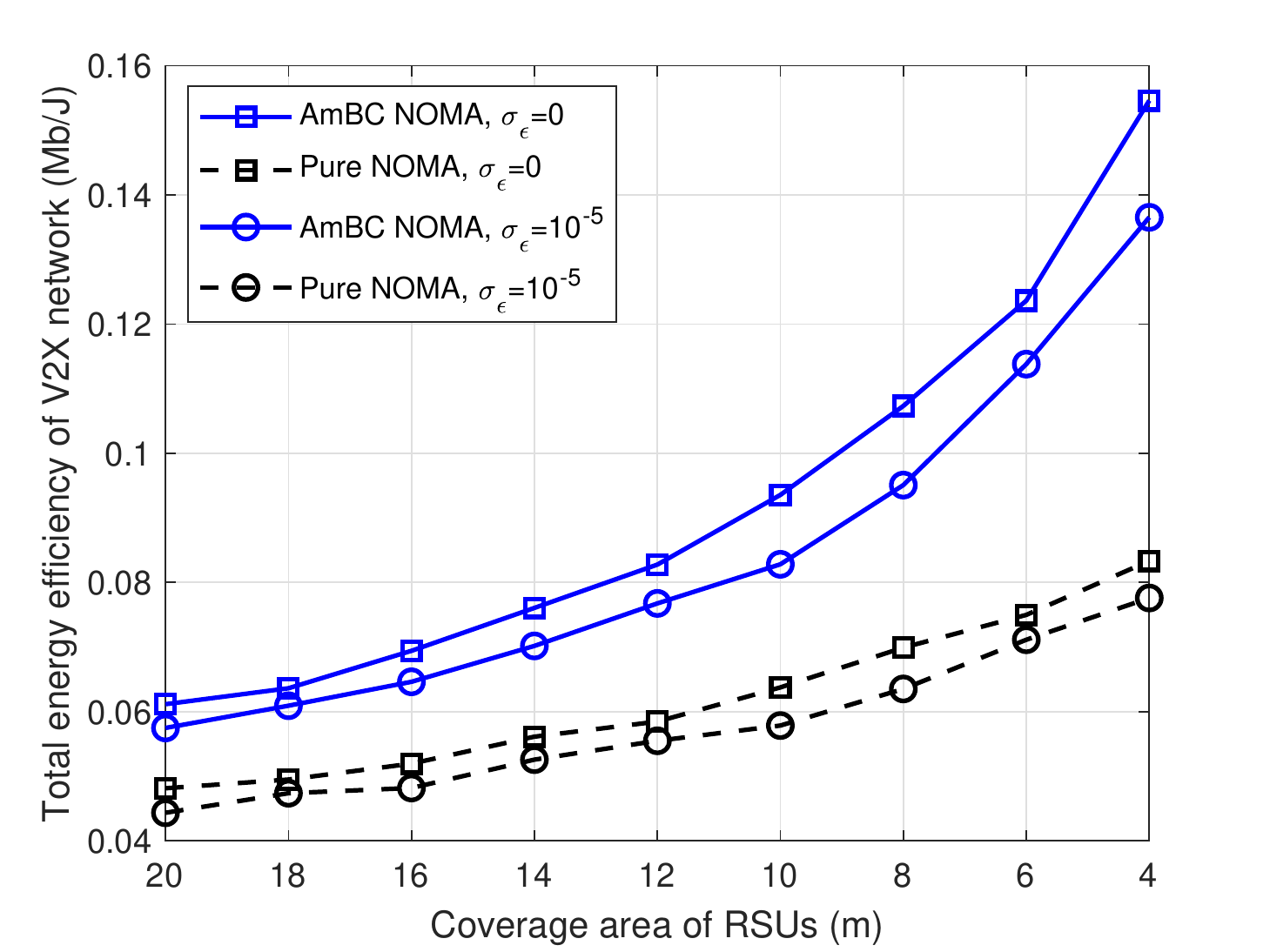}
    \caption{Comparison of system total energy efficiency versus coverage area of RSUs by varying $\sigma_{\epsilon}$.}
    \label{Fig5}
\end{figure}

Next, we show the impact of RSU coverage on the total energy efficiency of cooperative V2X communication for both AmBC enhanced NOMA and pure NOMA networks. Figure \ref{Fig5} plots the total energy efficiency versus the coverage area of RSUs. In this plot, we set the values of imperfect CSI as $\sigma_{\epsilon}=0,10^{-5}$. It can be observed that the achievable energy efficiency of both V2X networks is increasing with the decrease in the coverage area of RSUs. We can also see that the perfect CSI systems achieve higher energy efficiency than the system with imperfect CSI. However, in perfect and imperfect CSI cases, the proposed AmBC enhanced NOMA cooperative V2X communication significantly outperforms the conventional pure NOMA cooperative V2X communication. Another important thing that should also be noted is that the gap of total energy efficiency between the proposed network and the benchmark network increases as the coverage area of RSUs decreases, which shows the effectiveness of AmBC in short-range transmission.

\begin{figure}[t]
\centering
    \includegraphics[width=\columnwidth]{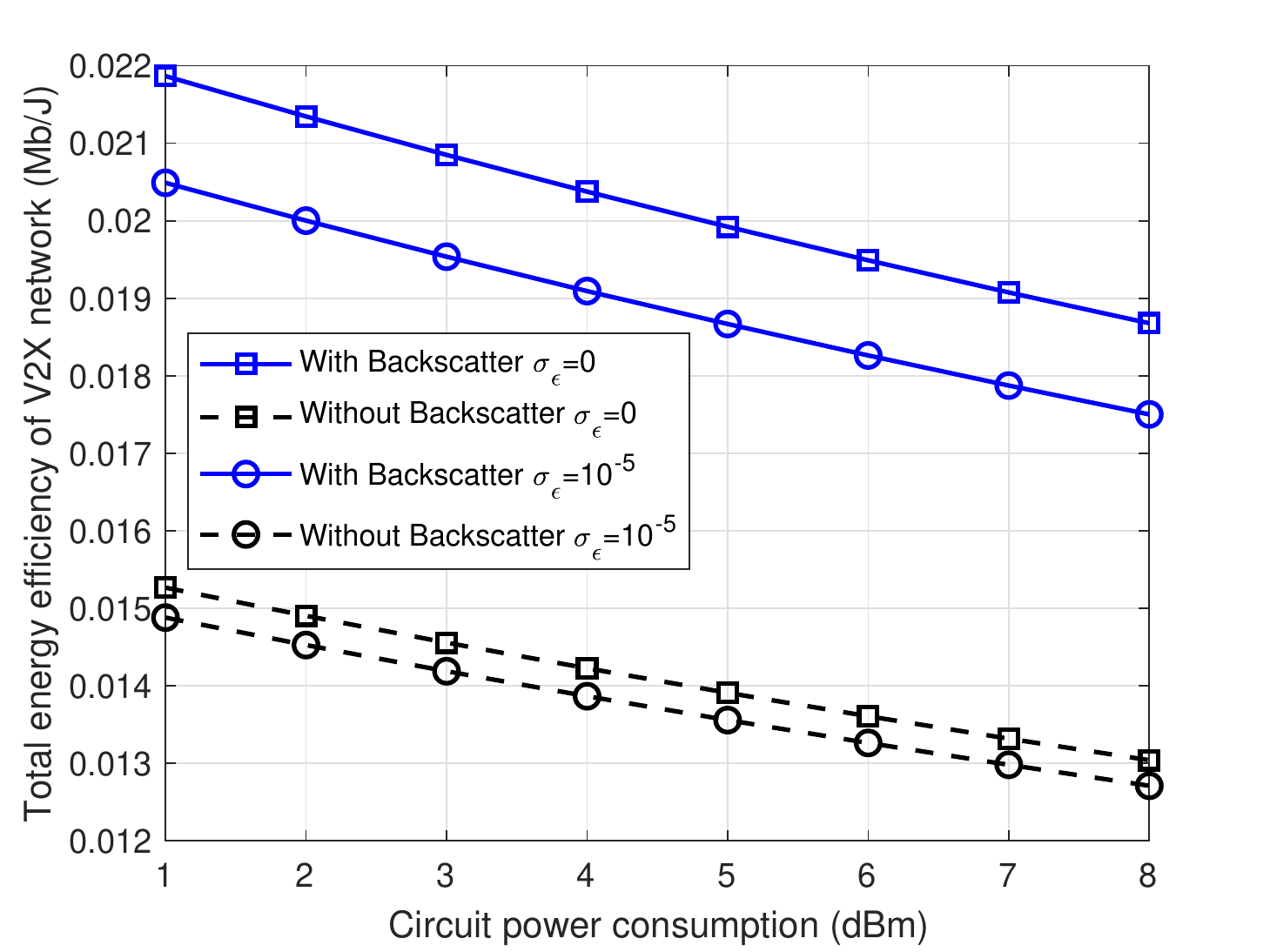}
    \caption{Comparison of system total energy efficiency versus system circuit power consumption by varying $\sigma_{\epsilon}$.}
    \label{Fig6}
\end{figure}

Last but not least, we check the impact of circuit consumption on the system's total energy efficiency for both networks. Figure \ref{Fig6} depicts the total energy efficiency versus the increasing values of circuit power consumption, where the values of $\sigma_{\epsilon}$ are 0 and $10^{-5}$. As expected, the achievable energy efficiency of both networks reduces as the circuit consumption increases. However, for both $\sigma_E=0$ and $\sigma_{\epsilon}=10^{-5}$, the proposed AmBC enhanced NOMA cooperative V2X communication achieves very high energy efficiency compared to the pure NOMA V2X communication. For instance, when $\sigma_{\epsilon}$ is $10^{-5}$ and the circuit power consumption is 2 dBm, the proposed AmBC NOMA V2X network achieves above 0.02 Mb/J. On the other side, the conventional pure NOMA V2X network can only achieve 0.0145 Mb/J approximately for similar system parameters. This shows the importance of the proposed system using AmBC in such scenarios.    

\section{Conclusion}
AmBC and NOMA are promising technologies for providing high spectral and energy efficiency in future Automotive-Industry 5.0 communications. This work has investigated a new optimization problem for enhancing energy efficiency in AmBC enhanced NOMA cooperative V2X communication. In particular, the transmit power of BS and RSUs and the reflection coefficient of backscatter sensors have been optimized simultaneously under perfect/imperfect CSI errors. The problem of total transmit power minimization has been formulated as non-convex, which was very hard and complex. The original problem has been transformed and decoupled into two sub-problems, and then the iterative sub-gradient method has been adopted to obtain an efficient solution. Comprehensive simulation results have also been provided and discussed to show the advantages of AmBC enhanced NOMA cooperative V2X communication against the benchmark pure NOMA cooperative V2X communication. 

Our proposed optimization framework can be further extended in multiple ways. For example, the considered system can be investigated under decode-and-forward and amplify-and-forward relaying protocols. The proposed system can also be studied when the RSUs operate in full-duplex such that the signal from BS to vehicles can be received in a one-time slot. Further, intelligent reflecting surfaces can be incorporated in the proposed model to replace the energy-constrained relays. These exciting yet explored topics will be done in the future.

\ifCLASSOPTIONcaptionsoff
  \newpage
\fi

\bibliographystyle{IEEEtran}
\bibliography{Wali_EE}

\end{document}